\documentclass[lettersize,journal]{IEEEtran}
\usepackage{amsmath,amsfonts}
\usepackage{algorithmic}
\usepackage{algorithm}
\usepackage{array}
\usepackage[caption=false,font=normalsize,labelfont=sf,textfont=sf]{subfig}
\usepackage{textcomp}
\usepackage{stfloats}
\usepackage{url}
\usepackage{verbatim}
\usepackage{graphicx}
\usepackage{cite}
\usepackage{subfig}
\usepackage{makecell}
\begin{document}

\title{Wi-Fi-based Personnel Identity Recognition: Addressing Dataset Imbalance with C-DDPMs}

\author{Jichen~Bian,~\IEEEmembership{Graduate Student Member,~IEEE,}
Chong~Tan,~\IEEEmembership{Member,~IEEE,}
Peiyao~Tang,
and Min~Zheng
\thanks{This work has been submitted to the IEEE for possible publication. Copyright may be transferred without notice, after which this version may no longer be accessible.}
\thanks{Jichen~Bian and Peiyao~Tang are with the Shanghai Insititute of Microsystem and Information Technology, Chinese Academy of Sciences, Shanghai 200050, China, and also with the University of Chinese Academy of Sciences, Beijing 100049, China (e-mail: jichen.bian@mail.sim.ac.cn; peiyao.tang@mail.sim.ac.cn).}
\thanks{Chong~Tan and Min~Zheng are with the Shanghai Insititute of Microsystem and Information Technology, Chinese Academy of Sciences, Shanghai 200050, China (e-mail: chong.tan@mail.sim.ac.cn; min.zheng@mail.sim.ac.cn).}
}

\markboth{Journal of \LaTeX\ Class Files,~Vol.~XX, No.~X, XXX~202X}%
{Shell \MakeLowercase{\textit{et al.}}: Wi-Fi-based Personnel Identity Recognition: Addressing Dataset Imbalance with C-DDPMs}

\IEEEpubid{0000--0000/00\$00.00~\copyright~2021 IEEE}

\maketitle

\begin{abstract}
Wireless sensing technologies become increasingly prevalent due to the ubiquitous nature of wireless signals and their inherent privacy-friendly characteristics. Device-free personnel identity recognition, a prevalent application in wireless sensing, is susceptibly challenged by imbalanced channel state information (CSI) datasets. This letter proposes a novel method for CSI dataset augmentation that employs Conditional Denoising Diffusion Probabilistic Models (C-DDPMs) to generate additional samples that address class imbalance issues. The augmentation markedly improves classification accuracies on our homemade dataset, elevating all classes to above 94\%. Post-review, we will release the dataset on IEEE DataPort and the code on GitHub.
\end{abstract}

\begin{IEEEkeywords}
Wi-Fi, channel state information, personnel identity recognition, diffusion model, vision transformer
\end{IEEEkeywords}

\section{Introduction}
\IEEEPARstart{P}{ersonnel} identity recognition (PIR) has emerged as a critical technology in various fields including intrusion detection, intelligent surveillance, and security management. Traditional methods, such as vision-based systems and sensor-based technologies \cite{9552382, 10190332}, are effective in specific scenarios but exhibit inherent limitations. Cameras often struggle to capture useful information under conditions with poor lighting or non-line of sight (NLOS) scenarios. Additionally, the extensive use of cameras can potentially lead to privacy concerns. Sensor-based techniques require individuals to carry specific ID devices, presenting inconveniences and a risk of loss. Addressing these challenges, the emergence of wireless signal-based identity recognition, especially using CSI data, has gained significant attention. CSI-based recognition eliminates the dependence on external light sources and does not necessitate any special devices to be carried by individuals.

In the field of CSI-based identity recognition, numerous studies have been conducted. WiWho \cite{7460727} is the first to introduce the method of gait-based person identification using Wi-Fi CSI. The study requires participants to walk in a straight line within a defined area for CSI collection and employs machine learning for classification. WiDFF-ID \cite{9953957} utilizes Intel 5300 wireless NIC for identifying individuals with CSI data and proposes a novel algorithm that combines deep convolutional neural network (DCNN) with principal component analysis (PCA). \cite{10065584} employs transfer learning (TL) for gait-based person identification, achieving over 98.6\% accuracy with less than four minutes of training per person in varied environments. Our previous work \cite{SimpleViTFi} presents SimpleViTFi, a lightful vision transformer (ViT) model, achieving high accuracy in Wi-Fi-based person identification with an IEEE 802.11ac public dataset \cite{10144501}, with superior efficiency and minimal computational demands.

Despite these advancements, CSI-based identity recognition systems still confront several challenges as follows: the prevalent reliance on the Intel 5300 NIC, which offers limited subcarriers and bandwidth due to its dated technology; a common oversight regarding the distribution of samples, which may lead to data imbalance across different classes.

In this letter, we propose an innovative method for CSI generation. Initially, we use the Asus RT-AC86U routers for CSI collection. Then, we convert the processed amplitude and phase of the CSI data into multi-channel CSI images. Subsequently, we utilize C-DDPMs to generate a substantial volume of new CSI samples. These synthesized samples are integrated with the imbalanced dataset to enrich the training process. By retraining our classification models with this augmented dataset, we observe notable improvements in classification accuracy specifically in classes with insufficient data. The augmented dataset demonstrates robust performance, indicative of the potential for C-DDPMs in addressing dataset imbalances within the field of CSI-based PIR. The contributions of this letter are summarized as follows:

\begin{enumerate}
    \item Construct a dataset comprising CSI data of 20 participants moving randomly within a meeting room, transforming the CSI data into 4-channel images consisting of amplitude and phase information after preprocessing.
    \item Apply C-DDPMs to generate realistic multi-channel CSI images, bridging the gap in data balance.
    \item Employ a light ViT model for classification training, getting commendable results with imbalanced sample distribution.
\end{enumerate}

\IEEEpubidadjcol

\section{Methodologies}

\subsection{Phase Error Correction}

Previous study \cite{8259000} points out the following sources of phase errors in CSI measurements due to hardware imperfection, including packet detection delay (PDD), sampling frequency offset (SFO), carrier frequency offset (CFO), random initial phase offset, and phase ambiguity. To mitigate these errors and obtain a more accurate representation of the phase information, a two-step phase correction process is employed.

The phase data is unwrapped along the subcarrier dimension, as shown in Equation~\ref{unwrapped}:

\vspace{-10pt}
\begin{equation}
    \hat{\phi}_{m+1} = \begin{cases} 
    \hat{\phi}_{m+1} - 2\pi & \text{if } \hat{\phi}_{m+1} - \hat{\phi}_m \geq \pi, \\
    \hat{\phi}_{m+1} + 2\pi & \text{if } \hat{\phi}_{m+1} - \hat{\phi}_m \leq -\pi, \\
    \hat{\phi}_{m+1} & \text{otherwise},
    \end{cases}
    \label{unwrapped}
\end{equation}

where $\hat{\phi}_k$ is the phase of the $k$-th subcarrier. The unwrapped phase is then fitted with a linear function to correct the phase error, as follows:

\vspace{-10pt}
\begin{equation}
    a = \frac{\hat{\phi}_n - \hat{\phi}_1}{k_n - k_1} - \frac{2\pi\delta}{N}, \\
\end{equation}

\vspace{-5pt}
\begin{equation}
    b = \frac{1}{n} \sum_{j=1}^{n} \hat{\phi}_j - \frac{2\pi\delta}{nN} \sum_{j=1}^{n} k_j + \beta,    
\end{equation}

\vspace{-5pt}
\begin{equation}
    \tilde{\phi}_j = \hat{\phi}_j - (ak_j + b),    
\end{equation}

where \(k_j\) is the index of the \(j\)-th subcarrier. The real phase can be calculated by subtracting the linear term \(ak_j+b\) from the measured \(\hat{\phi}_j\).

\begin{figure}[htbp]
\centering
{\includegraphics[width=0.32\linewidth]{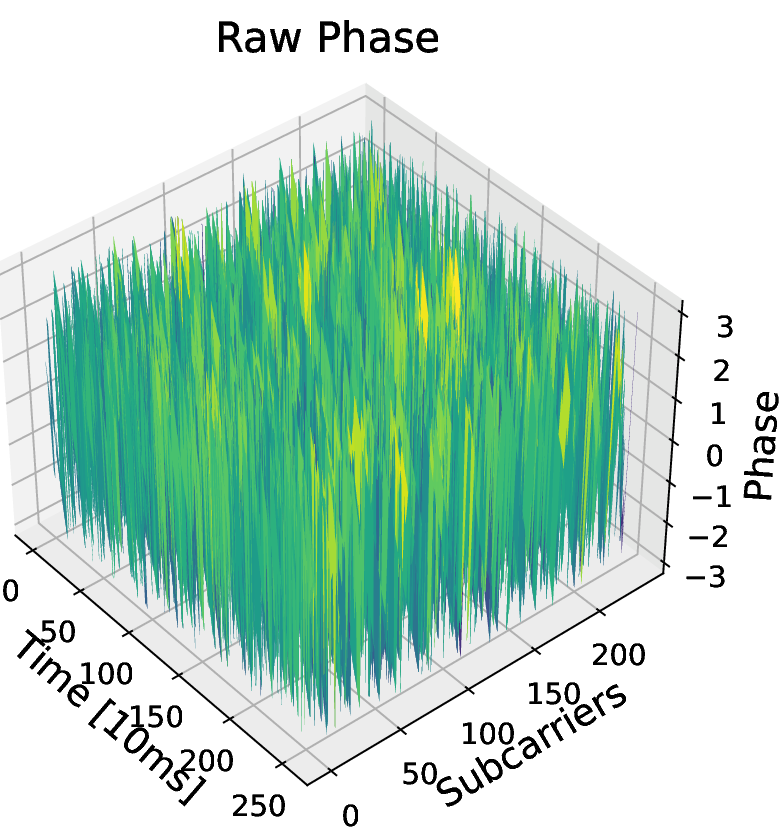}}
{\includegraphics[width=0.32\linewidth]{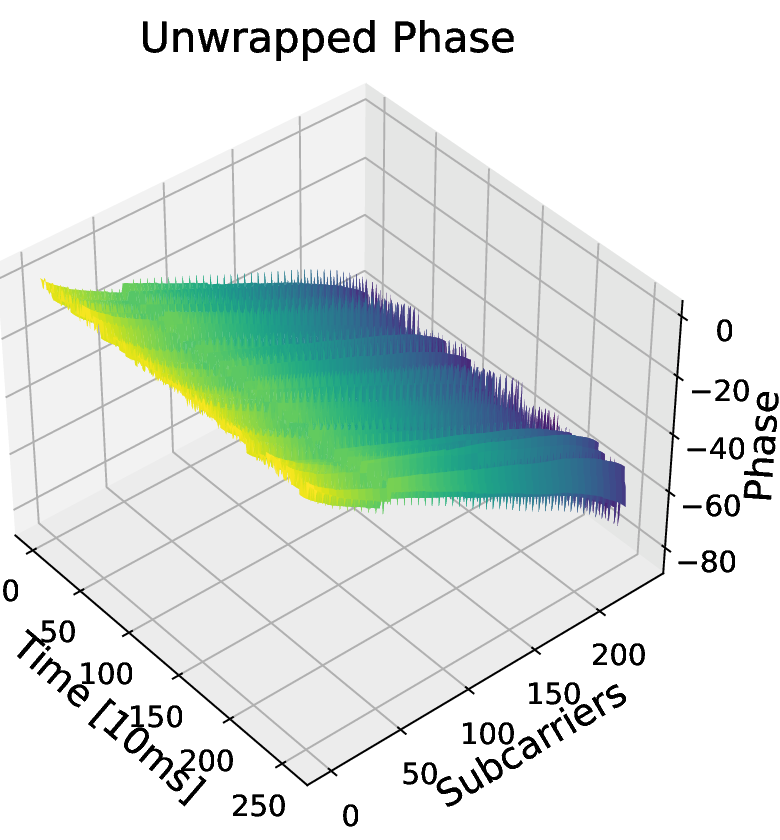}}
{\includegraphics[width=0.32\linewidth]{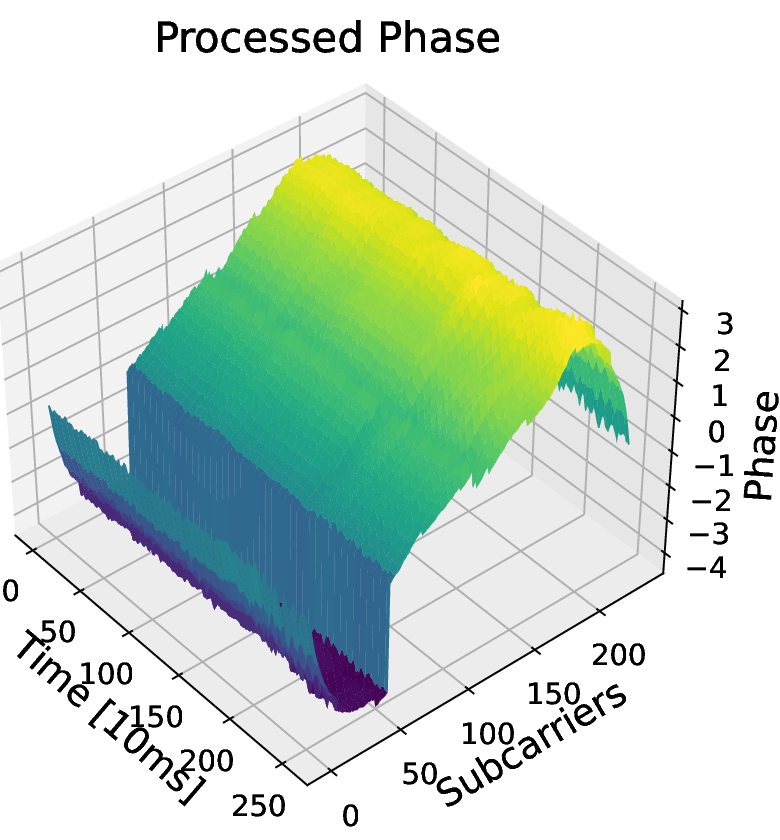}}
\caption{Phase error processing. From left to right: The raw phase data exhibit noise and discontinuities. The middle figure shows the unwrapped phase, correcting discontinuities. The processed phase on the right demonstrates the linear fitting of the phase error.}
\label{phase-error}
\end{figure}

\vspace{-10pt}
\subsection{Conditional Denoising Diffusion Probabilistic Models (C-DDPMs)}

DDPMs provide a unique approach to generative modeling, integrating diffusion processes with deep learning techniques to enhance data diversity. Building upon this foundation, C-DDPMs further extend the capabilities of DDPMs by embedding vectorized labels, enabling the generation of data across multiple classes \cite{9711284}. This inclusion of label-specific information allows C-DDPMs to effectively balance the distribution of CSI datasets. The main operation of C-DDPMs involves both the forward (diffusion) and reverse (denoising) processes:

\textit{Forward Process}: The forward process in C-DDPM is a Markov chain that incrementally adds Gaussian noise to the data. This process is mathematically represented as:
\begin{equation}
q(x_t | x_{t-1}) = \mathcal{N}(x_t; \sqrt{1 - \beta_t}x_{t-1}, \beta_t \mathbf{I}),
\end{equation}
where $\left\{\beta_t\right\}^T_{t=1}$ are the variances that linearly increase with each time step \( t \). In this letter, we define \( T = 500 \) and $\left\{\beta_t\right\}^T_{t=1}$ increase from $10^{-4}$ to $0.28$. This process gradually transforms the data into an isotropic Gaussian noise-like state.
\textit{Reverse Process}: The objective is to progressively eliminate the noise, thereby reconstructing the original data distribution. It is challenging to obtain a probability distribution for the entire dataset; hence, it necessitates developing a parameterized model to learn the reverse process. It is defined as:
\begin{equation}
p_\theta(x_{t-1} | x_t, c) = \mathcal{N}(x_{t-1}; \mu_\theta(x_t, t, c), \Sigma_\theta(x_t, t, c)),
\end{equation}
\vspace{-10pt}
\begin{equation}
\mu_\theta(x_t, t, c) = \frac{1}{\sqrt{\alpha_t}}(x_t-\frac{1-\alpha_t}{\sqrt{1-\bar{\alpha}_t}}\epsilon_\theta(x_t,t,c)),
\end{equation}
where $\mu_\theta(x_t, t, c)$ represents the mean function in the reverse process; $x_t$, $\alpha_t$ and $c$ (vectorized label) are known variables, and $\bar{\alpha}_t$ is the cumulative product of \( \alpha_t \) up to time step \( t \). The variability in $\mu_\theta$ is instead captured by the noise prediction term $\epsilon_\theta(x_t, t, c)$, which is a reparametrization that shifts the burden of modeling the change in $\mu_\theta$ onto $\epsilon_\theta$. \cite{ho2020denoising} simplifies the model by setting $\Sigma_\theta(x_t, t, c)$ as a quantify.
The training of DDPMs, as detailed in \cite{ho2020denoising}, involves using stochastic gradient descent to optimize the reverse process. The authors in \cite{ho2020denoising} propose the simplified loss function through theoretical derivation and experimental validation:
\begin{equation}
L_{simple}(\theta) = \mathbb{E}_{t, x_0, \epsilon_t, c}[\|\epsilon_t - \epsilon_\theta(\sqrt{\bar{\alpha}_t}x_0 + \sqrt{1-\bar{\alpha}_t}\epsilon_t t)\|^2]
\end{equation}
At this point, the task of predicting the target distribution is transformed into predicting the noise term $\epsilon_t$ at each step.

\begin{figure}[htbp] 
\centerline{\includegraphics[width=1.0\linewidth]{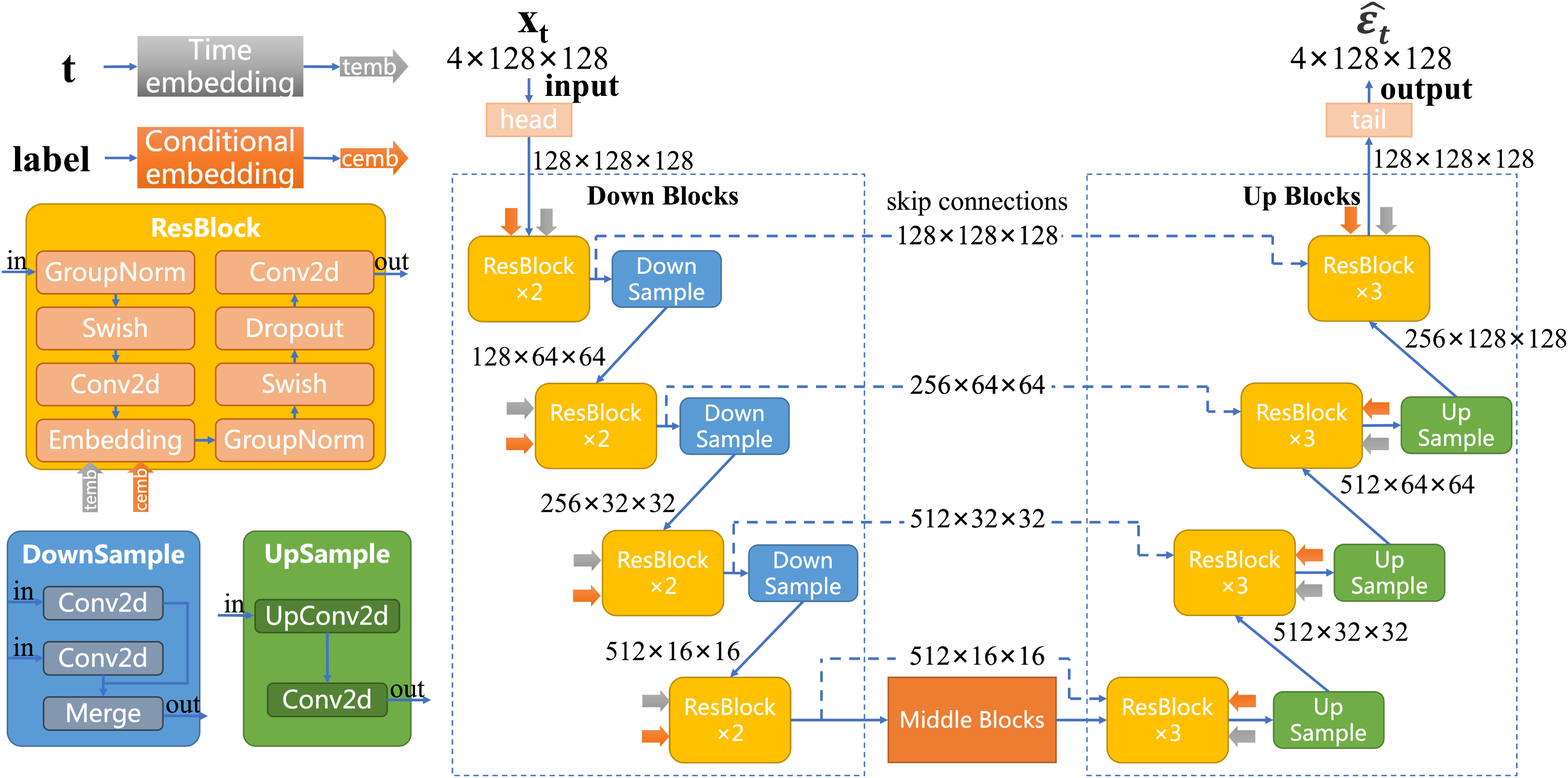}} 
\caption{The U-Net architecture of the C-DDPMs.} 
\label{Unet}
\end{figure}

The implementation framework of C-DDPMs is a U-Net architecture as shown in Fig~\ref{Unet}, which begins with a head block expanding the dimensionality of the input CSI images \(x_t\). This block is followed by a series of residual and down-sampling layers that condense the spatial dimensions and amplify the depth of feature representation. The middle block serves for deep feature extraction, and up-sampling layers then reconstruct the spatial dimensions. The tail block compresses the dimensions back, finalizing the output \(\hat{\epsilon_t}\). Time and label information are integrated into the model via embeddings, enabling precise noise prediction and class-specific data generation.

\vspace{-10pt}
\section{Experimental Results}

\subsection{Datasets}

We create a dataset by collecting CSI data in a 7 by 5 meters meeting room, furnished with a conference table, several chairs, and a locker. Two Asus RT-AC86U routers, equipped with nexmon tools \cite{10.1145/3349623.3355477}, are positioned within the room: one for periodic single-antenna data transmission every 10ms and another with four antennas for continuous data packet monitoring.

\begin{figure}[htbp]
\centering
\subfloat[Amp (empty room)]
{\includegraphics[width=0.5\linewidth]{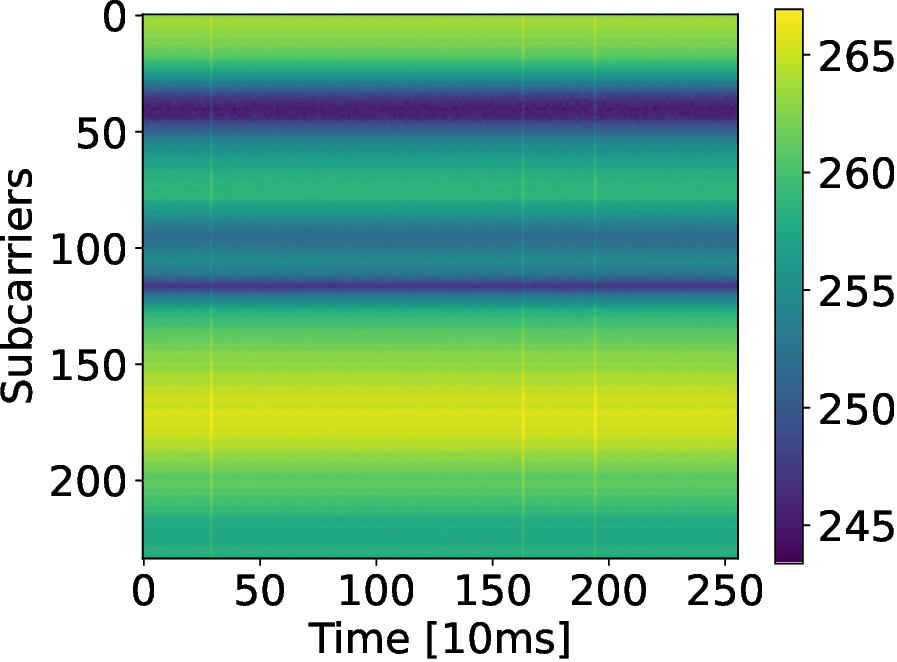}\label{vision0}}
\subfloat[Phase (empty room)]
{\includegraphics[width=0.49\linewidth]{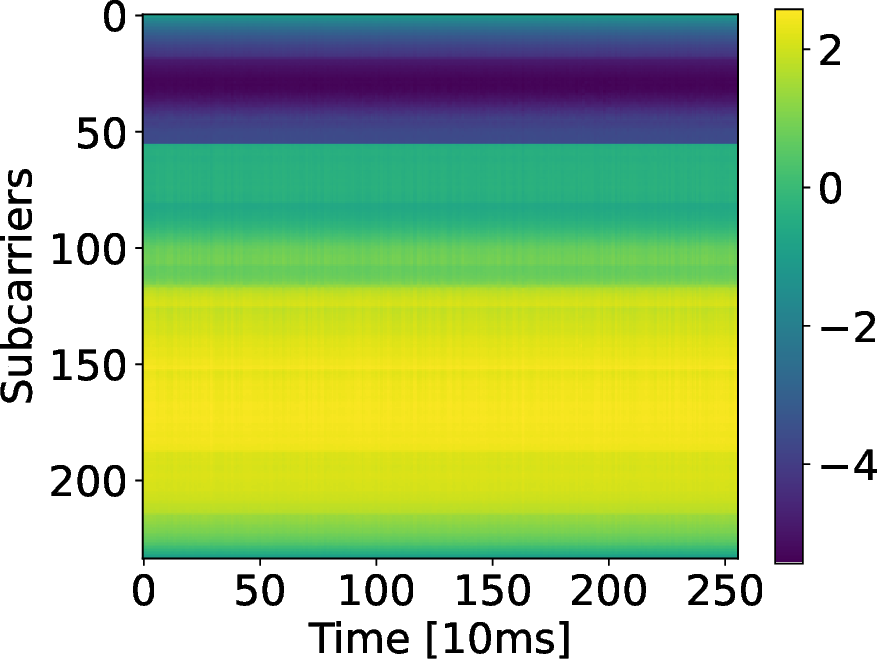}\label{vision0-phase}}\\
\vspace{-10pt}
\subfloat[Amp (participant 2)]
{\includegraphics[width=0.5\linewidth]{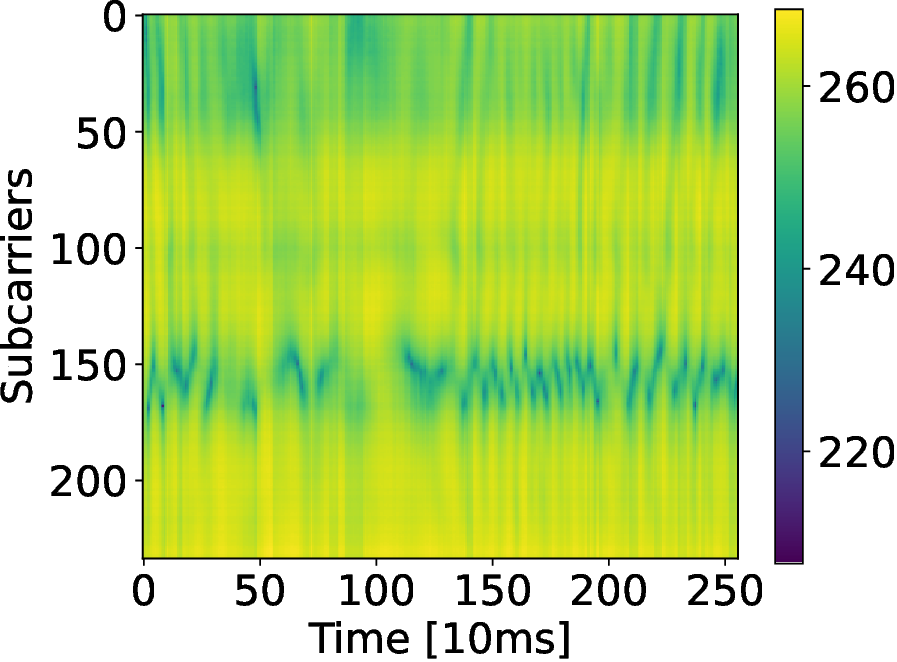}\label{vision2}}
\subfloat[Phase (participant 2)]
{\includegraphics[width=0.49\linewidth]{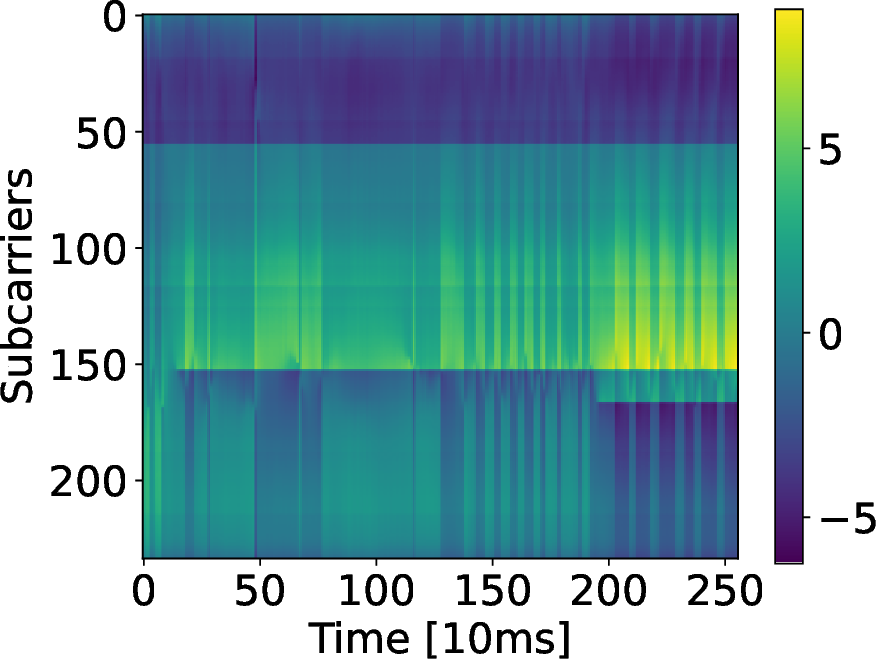}\label{vision2-phase}}
\caption{Amplitude (in dB scale) and phase of CSI data for an empty room (a)-(b) and the participant 2 (c)-(d) walking. Each trace is about 2.56 seconds long (\(x\)-axis) and shows the amplitude or the phase on each subcarrier (\(y\)-axis). Note that CSI is not available on the control carriers and pilot carriers, so there are 234 carriers left to use.}
\label{original-csi}
\end{figure}

The dataset encompasses CSI measurements, \( CSI_{c,k,t} \), from 20 distinct participants, each moving randomly within the room. These measurements capture dynamic changes in the wireless channel due to human movement, with a data tensor composed of \( C = 4 \) channels—comprising amplitude and processed phase values from one external and one internal antenna—across \( K = 256 \) subcarriers, and \( T_{slot} = 256 \) discrete time slots. In total, 21 such sets are assembled, including a baseline set with the empty room. The data fluctuations within about 2.56 seconds window, as illustrated in Fig~\ref{original-csi}, reflect distinct motion-induced CSI perturbations.

\vspace{-10pt}
\subsection{Experimental Setup}

\begin{figure}[htbp] 
\centerline{\includegraphics[width=1.0\linewidth]{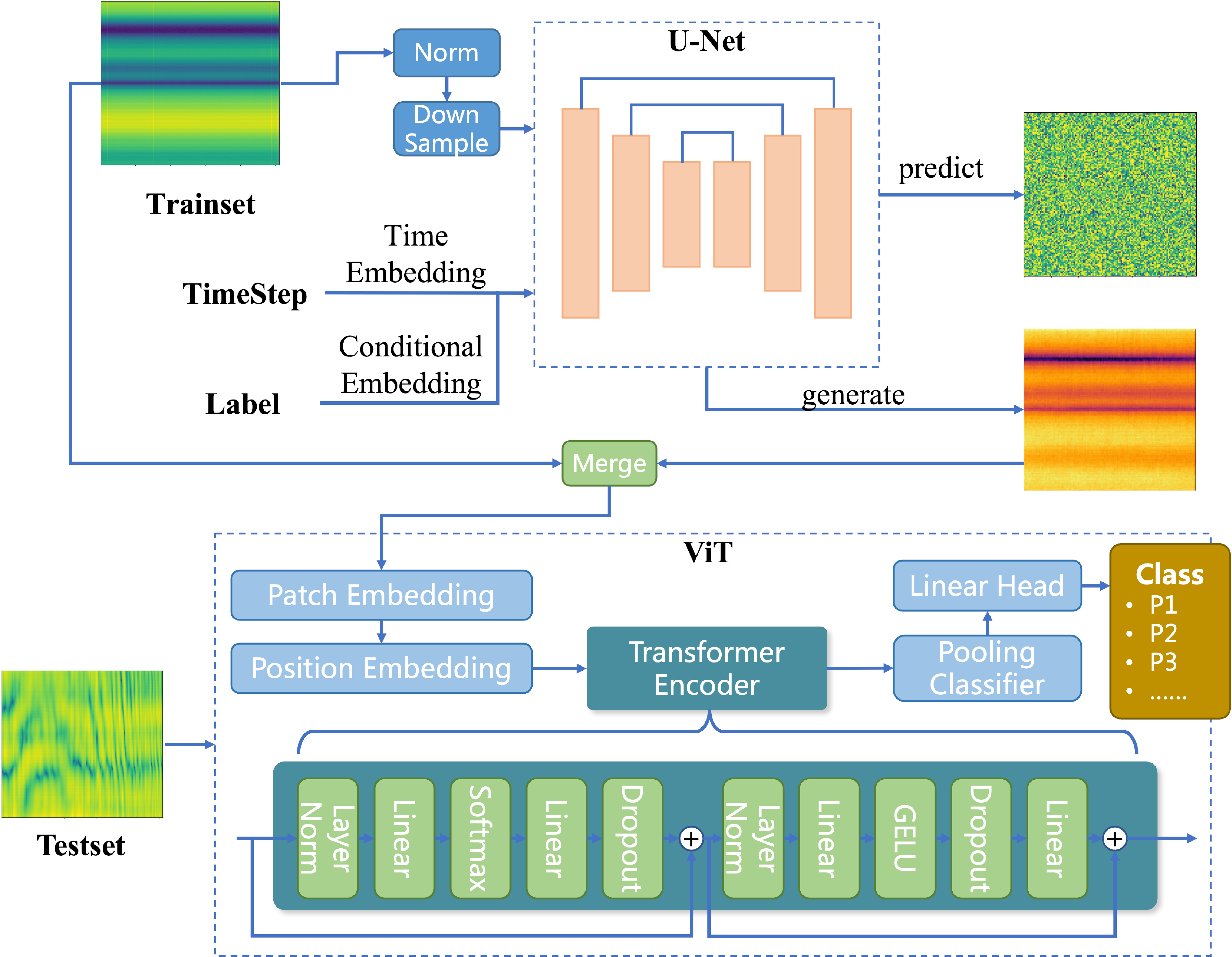}} 
\caption{The entire framework, incorporating U-Net for data generation and ViT for classification. It illustrates the workflow from generating synthetic CSI images to classifying them into predefined classes.} 
\label{framework}
\end{figure}

The experimental setup is designed to evaluate the efficacy of the C-DDPMs in augmenting CSI datasets and improving classification performance. Three scenarios are considered: the balanced dataset, the imbalanced dataset and the dataset augmented by C-DDPMs. The classification model, SimpleViTFi, is employed as proposed in our previous work \cite{SimpleViTFi}, which is a lightweight ViT model tailored for CSI data. All experiments share the same model architecture and hyperparameters as shown in Table~\ref{vit-params}, with the only difference being the training set. The entire framework including data generation and classification is shown in Fig~\ref{framework}.

\begin{table}[htbp]
\centering
\caption{Parameters of SimpleViTFi}
\begin{tabular}{l|l}
\hline
\textbf{Parameter} & \textbf{Value} \\
\hline
Patch Embedding & \makecell*[l]{downsampling: $4 \times 128 \times 128$ \\ 16 patches: $16 \times 4 \times 128 \times 8$} \\
Position Embedding & Learnable Embedding \\
Transformer Encoder & \makecell*[l]{dim: 64\\ depth: 2\\ heads: 4\\ mlp\_dim: 128\\ dropout\_rate: 0.15\\ learning\_rate: 1e-4\\ weight\_decay: 0.1\\ loss\_function: CrossEntropyLoss} \\
Pooling & average pooling \\
Classifier Head & $\mathbf{Y} = \mathbf{W}\left(\frac{\mathbf{Z} - \text{E}[\mathbf{Z}]}{\sqrt{\text{Var}[\mathbf{Z}] + \epsilon}}\right) + \mathbf{b}$ \\
\hline
\end{tabular}
\label{vit-params}
\end{table}

Initially, we test the SimpleViTFi on the balanced dataset where each class splits 73\% of the data for training and the remaining 27\% for testing. Under this dataset, SimpleViTFi achieves 99.72\% accuracy, demonstrating its effectiveness on normal CSI images. 

\begin{figure}[htbp]
\centering
\subfloat[Imbalanced Dataset]
{\includegraphics[width=0.5\linewidth]{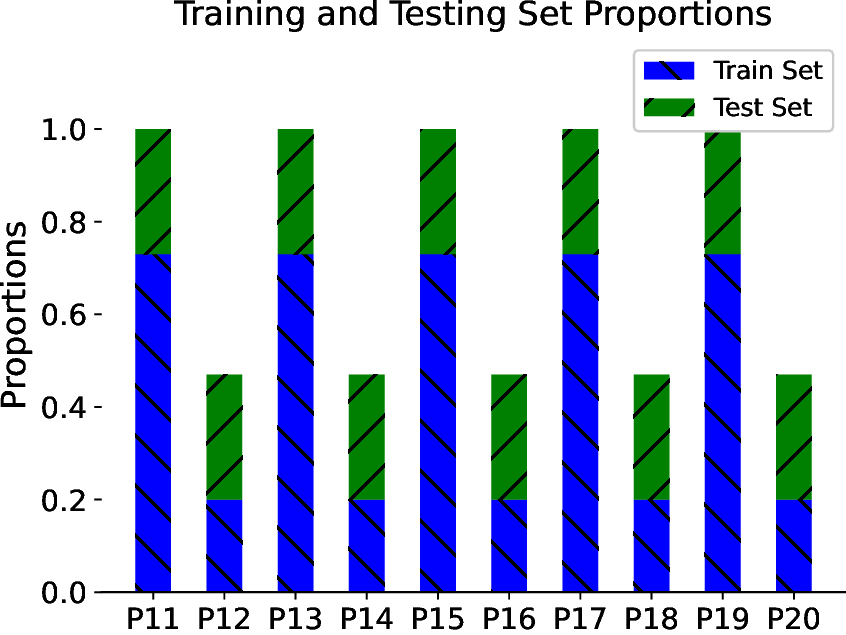}}
\subfloat[Augmented Dataset]
{\includegraphics[width=0.5\linewidth]{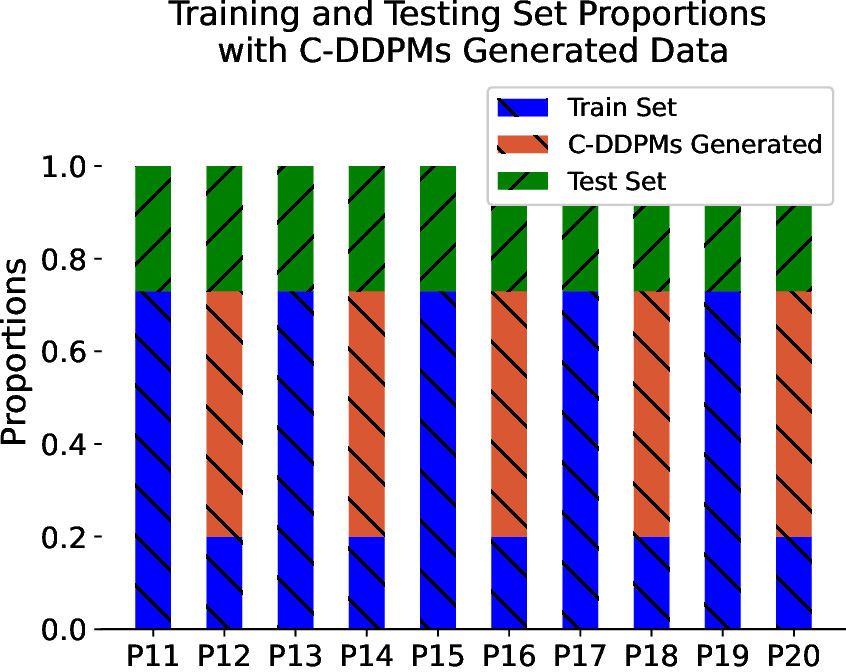}}\\
\caption{Class distribution in the training and testing sets.}
\label{dataset-proportions}
\end{figure}

While considering practical scenarios where the availability of data for certain classes may be limited, we then simulate an imbalanced dataset as shown in Fig~\ref{dataset-proportions}(a). For classes P11, P13, P15, P17, and P19, the proportions in the training set are reduced to only 20\%. Then, we employ C-DDPMs to train on the imbalanced training set, replenishing and balancing the training set as shown in Fig~\ref{dataset-proportions}(b). To showcase the effectiveness of the diffusion model in improving CSI-images classification accuracy, we train the SimpleViTFi model on each imbalanced dataset and augmented dataset.

\vspace{-10pt}
\subsection{Performance Evaluation}

\begin{figure}[htbp]
\centering
\subfloat[$t = 0$]
{\includegraphics[width=0.34\linewidth]{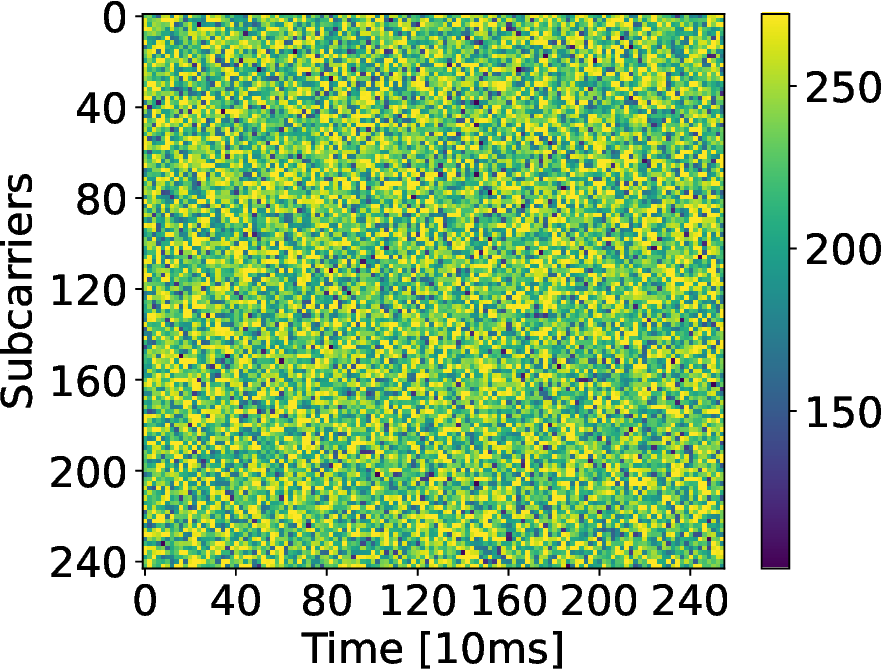}\label{step0}}
\subfloat[$t = 199$]
{\includegraphics[width=0.34\linewidth]{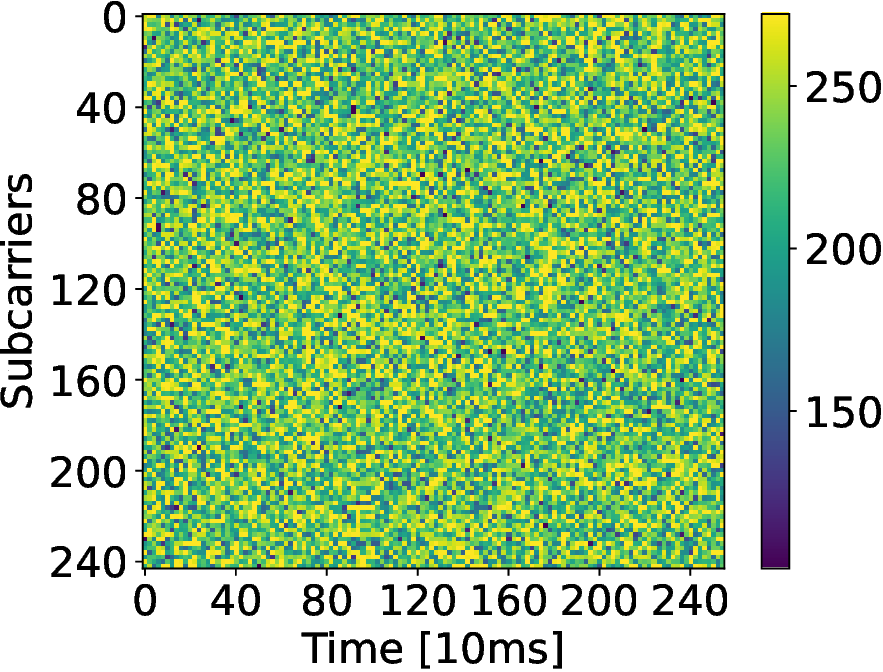}\label{step4}}
\subfloat[$t = 399$]
{\includegraphics[width=0.34\linewidth]{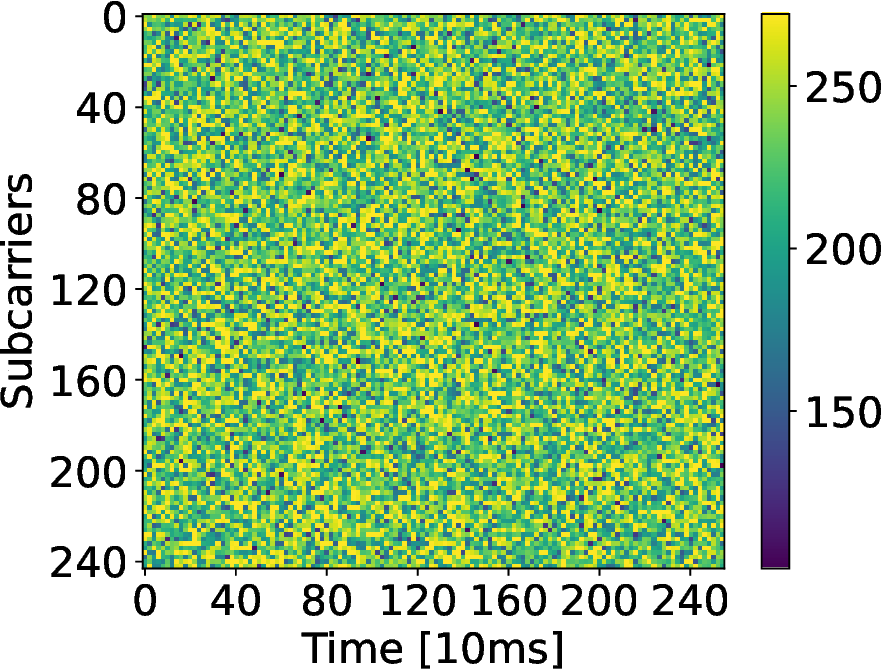}\label{step8}}\\
\vspace{-12pt}
\subfloat[$t = 429$]
{\includegraphics[width=0.34\linewidth]{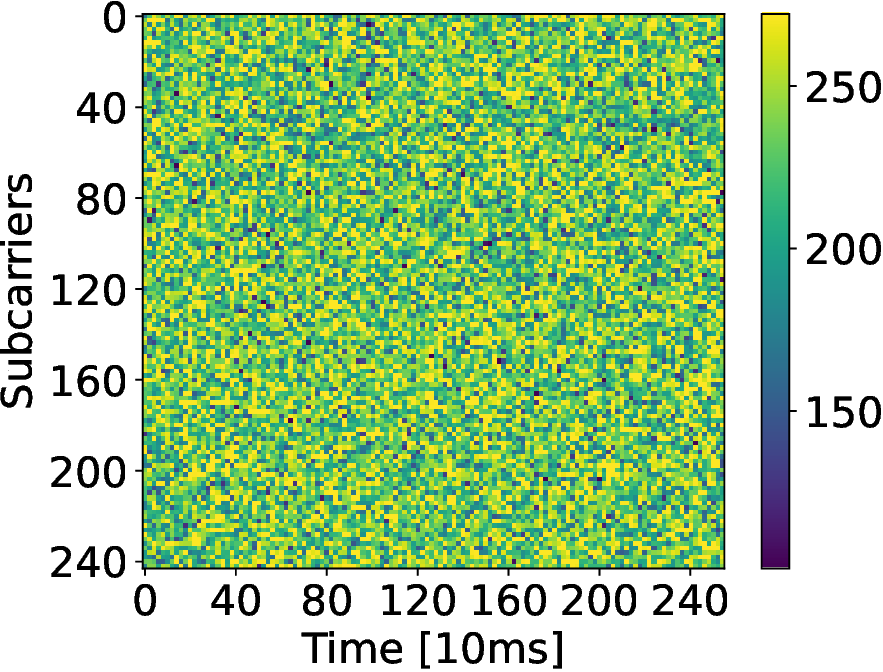}\label{step11}}
\subfloat[$t = 449$]
{\includegraphics[width=0.34\linewidth]{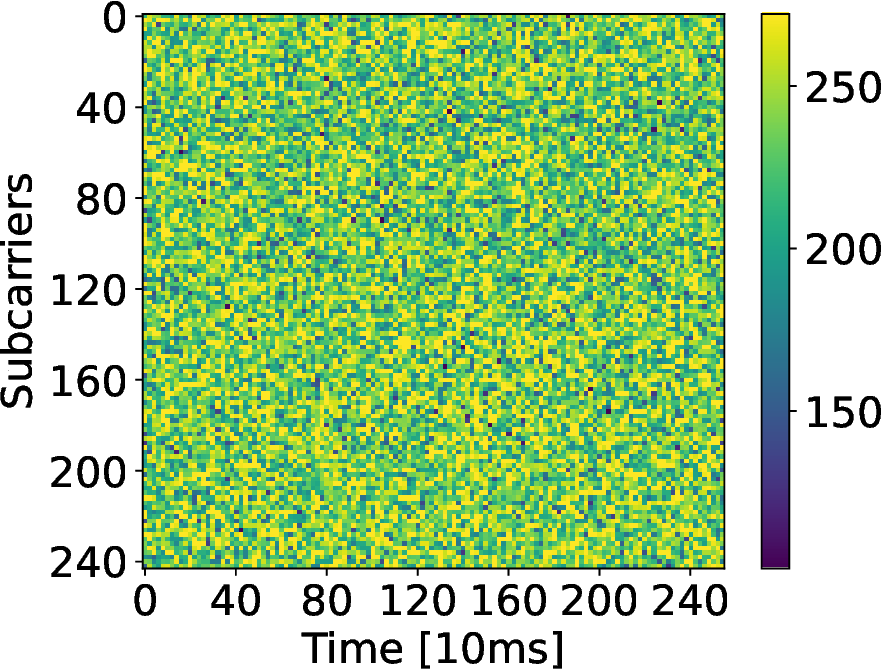}\label{step13}}
\subfloat[$t = 469$]
{\includegraphics[width=0.34\linewidth]{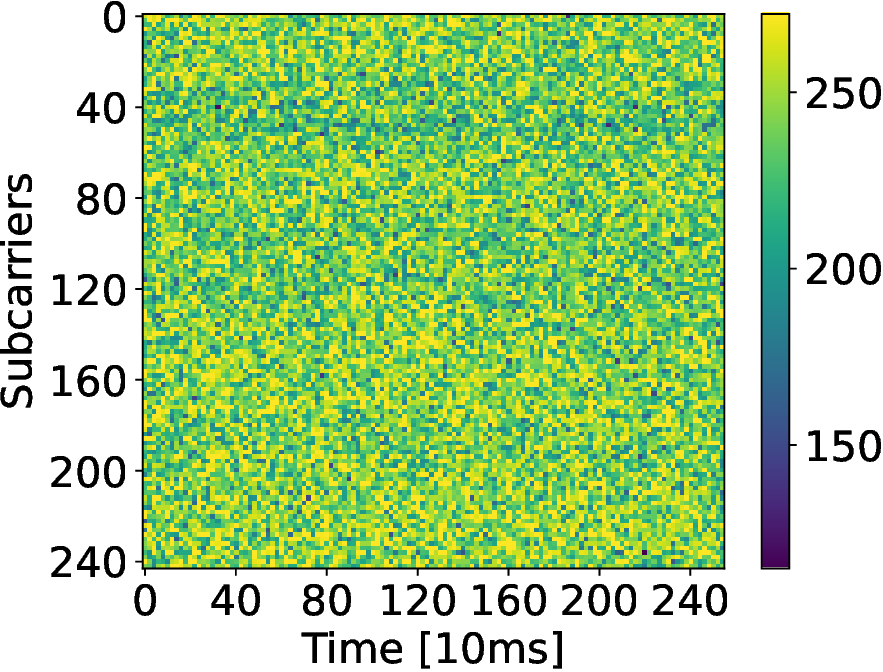}\label{step15}}\\
\vspace{-12pt}
\subfloat[$t = 479$]
{\includegraphics[width=0.34\linewidth]{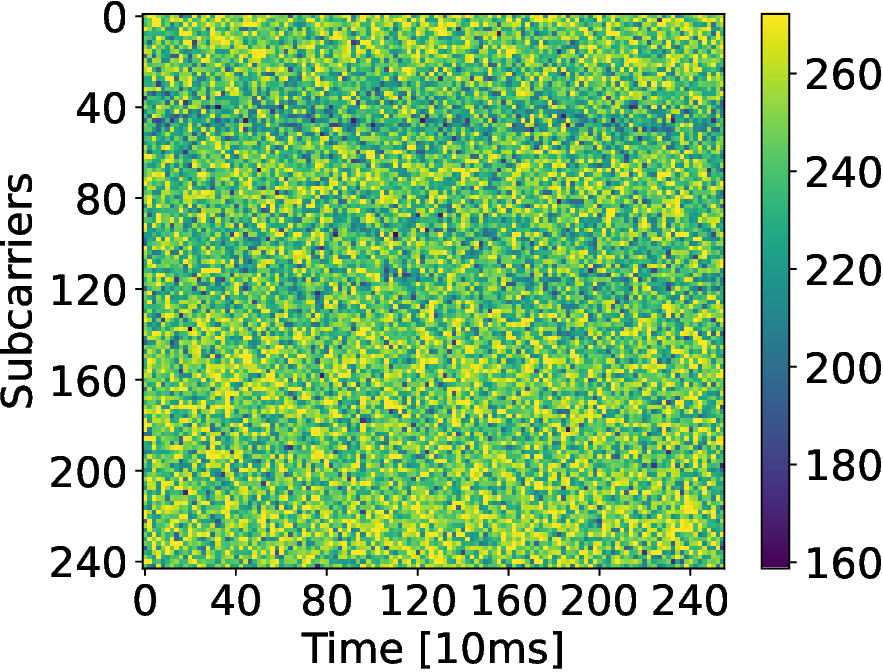}\label{step16}}
\subfloat[$t = 489$]
{\includegraphics[width=0.34\linewidth]{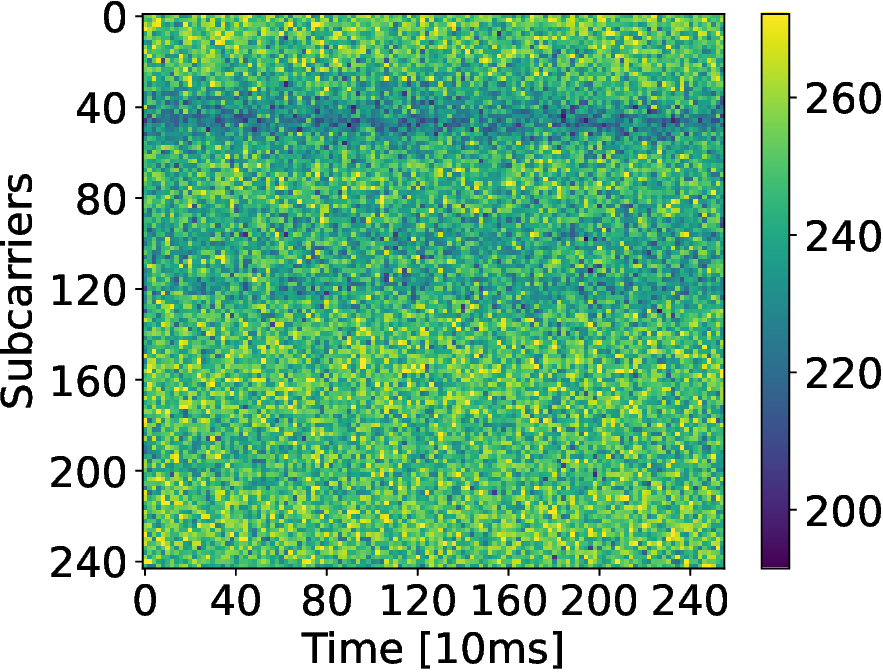}\label{step17}}
\subfloat[$t = 499$]
{\includegraphics[width=0.34\linewidth]{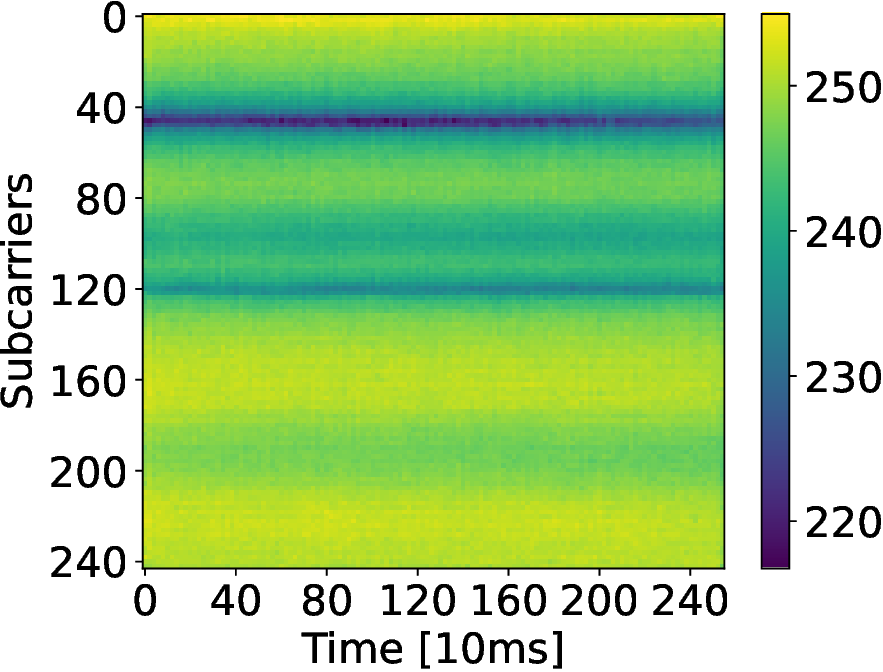}\label{step18}}
\caption{Visualization of the C-DDPM-generated process with time step \( t \) indicating the current step of the temporary result generated. The process illustrates the progressive denoising of sampled Gaussian noise and gradually transforming into the simulated CSI amplitude image.}
\label{generating-process}
\end{figure}

Fig~\ref{generating-process} visualizes the process of transforming Gaussian noise into synthetic CSI images with C-DDPMs. This step-by-step generation from noise to a high-fidelity imitation of the CSI data is crucial for augmenting the training set without introducing data bias. The denoising process is represented as:

\vspace{-10pt}
\begin{equation}
x_{t-1} = \tilde{\mu_t} + \tilde{\beta_t}^2 \epsilon,
\end{equation}
\vspace{-25pt}

\begin{equation}
\tilde{\mu_t} = \sqrt{\frac{1}{\alpha_t}} \left( x_t - \frac{\beta_t}{\sqrt{1 - \bar{\alpha}_t}} \hat{\epsilon} (x_t, t, c) \right),
\end{equation}
\vspace{-20pt}

\begin{equation}
\tilde{\beta_t}^2 = \frac{1 - \bar{\alpha}_{t-1}}{1 - \bar{\alpha}_t} \beta_t,
\end{equation}

where \(\hat{\epsilon}\) represents the noise predicted by C-DDPMs at time step \(t\) given the label \(c\). \(\hat{\epsilon}\) facilitates the calculation of the mean \(\tilde{\mu_t}\), which is then used to denoise \( x_t \) and estimate \( x_{t-1} \).

\begin{figure}[htbp] 
\centerline{\includegraphics[width=1.0\linewidth]{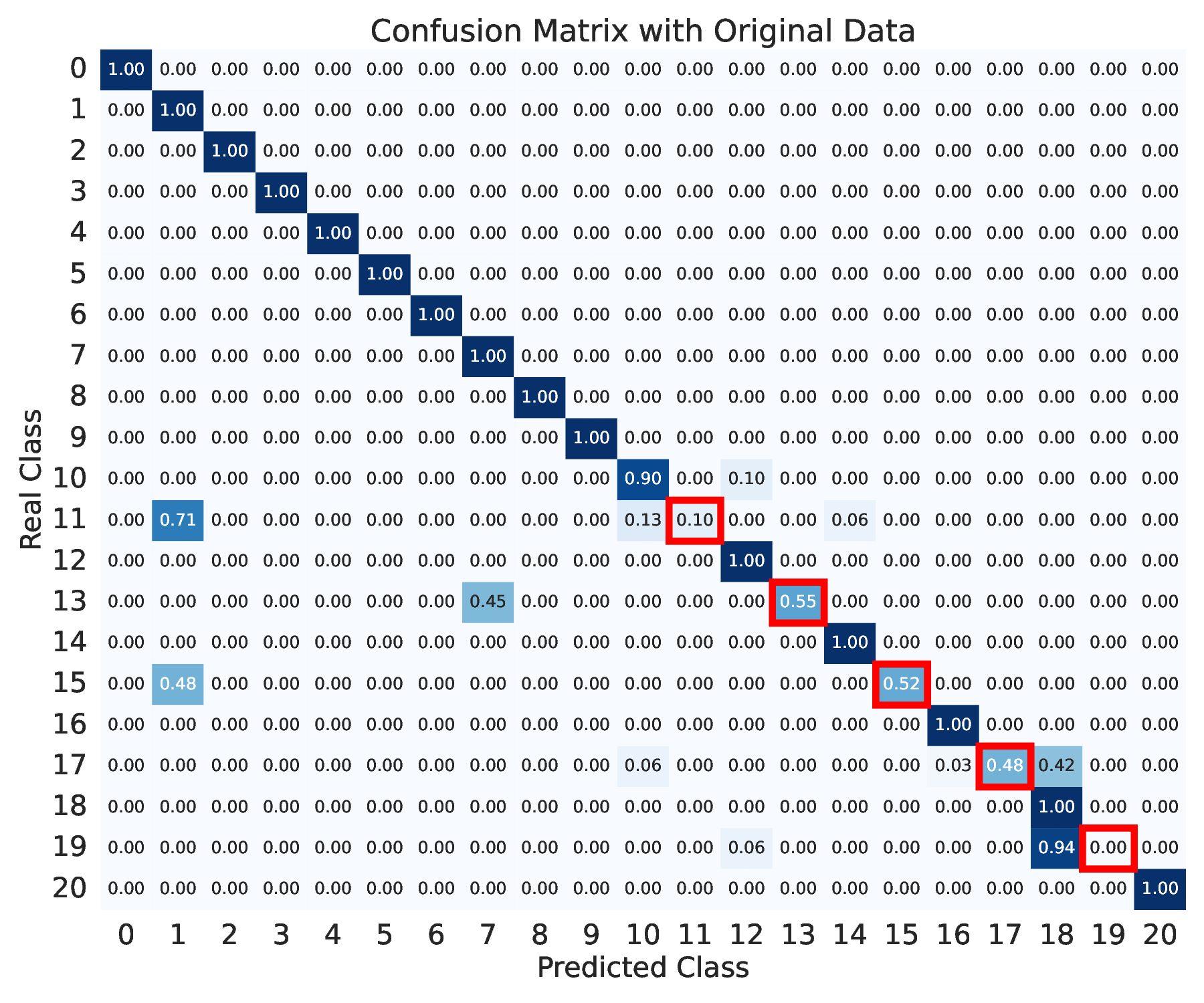}} 
\caption{Confusion Matrix of model performance on the testing set using only the original CSI training data. The matrix shows the accuracy per class with diagonal elements indicating the percentage of correct classifications. Classes encased in red boxes represent classes with a smaller number of data samples, which correspond to lower classification accuracies.}
\label{confusion-matrix-raw}
\end{figure}

\begin{figure}[htbp] 
\centerline{\includegraphics[width=1.0\linewidth]{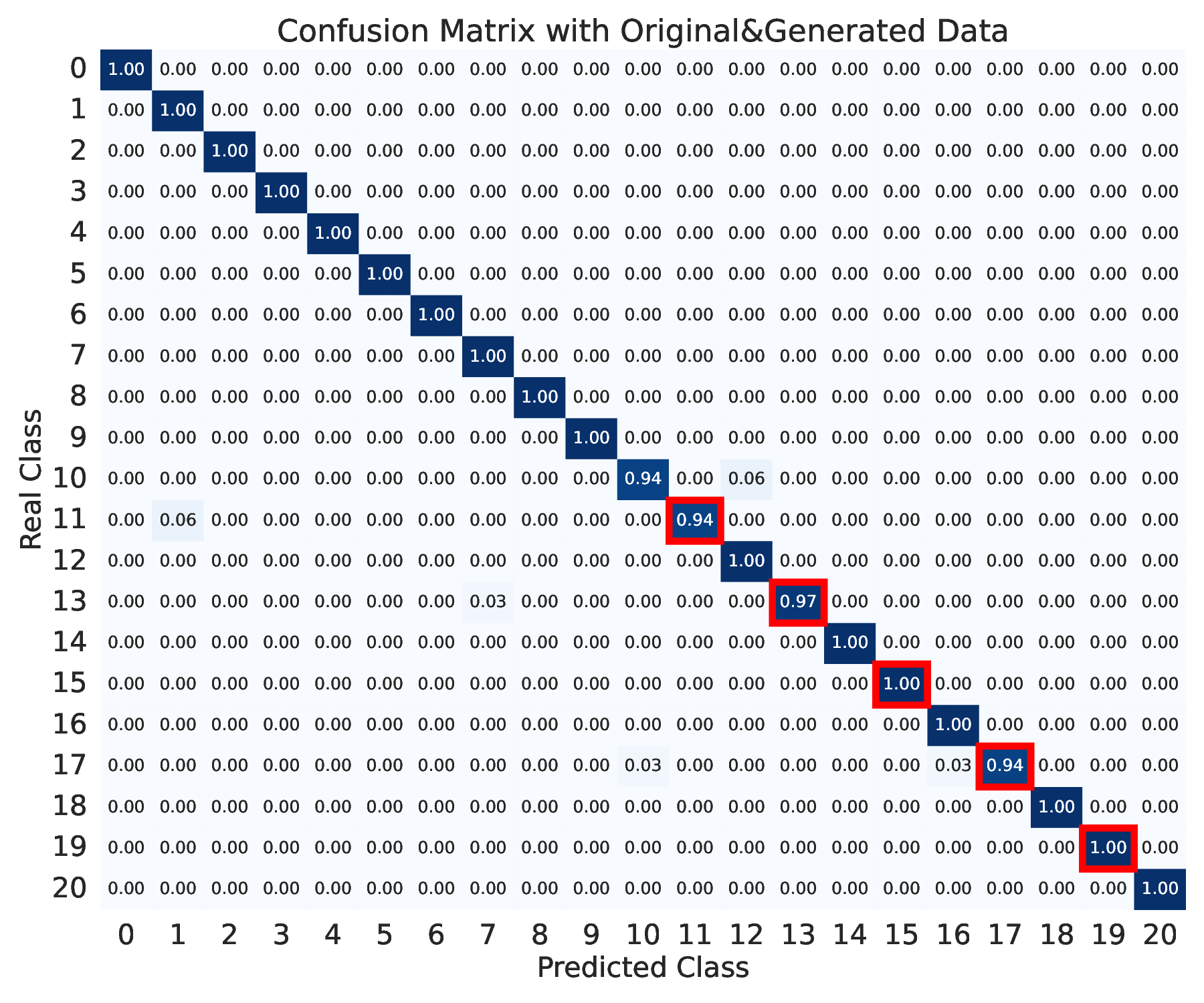}} 
\caption{Confusion Matrix showcasing the classification performance of the model trained on a dataset augmented with C-DDPMs. The red boxes show significant improvement in classification accuracy after the inclusion of generated data.}
\label{confusion-matrix-diffusion}
\end{figure}

Fig~\ref{confusion-matrix-raw} and Fig~\ref{confusion-matrix-diffusion} illustrate the confusion matrices of the ViT model's performance before and after the data augmentation, respectively. In Fig~\ref{confusion-matrix-raw}, the confusion matrix with the original data showcases disparities in accuracy across different classes, with diagonal elements indicating the percentage of correct classifications. Classes that are boxed in red indicate those with fewer data samples and, therefore, lower accuracies.

Fig~\ref{confusion-matrix-diffusion} displays the confusion matrix after including generated data into the training set. The red boxes in this matrix highlight the classes that were previously underrepresented and have now achieved a significant boost in accuracy, with each class achieving at least 94\% accuracy. 

These results validate that augmenting training datasets with high-quality generated data can be a potent strategy to enhance the robustness of classification models. On the one hand, the ability of C-DDPMs to model the data distribution ensures that the generated samples closely approximate the true distribution, which are not only replicas of existing data but also diverse and expanding samples. On the other hand, the enriched dataset provides a more comprehensive view of the data space, mitigating the risk of overfitting and leading to improved model generalization on unseen data.

\section{Conclusion}

In this letter, we introduce a novel data augmentation technique for device-free PIR with C-DDPMs to balance multi-class CSI datasets, significantly enhancing the classification accuracy of a lightweight ViT model amidst the challenges of class imbalance. As validated by sets of experimental results, the proposed technique demonstrates remarkable enhancement in the diversity and balance of the datasets, leading to improved model robustness and reliability.

\clearpage
\bibliographystyle{IEEEtran}
\bibliography{reference.bib}

\end{document}